# Misrepair mechanism in the development of atherosclerotic plaques

Jicun Wang-Michelitsch[1]*, Thomas M Michelitsch[2]

[1]Department of Medicine, Addenbrooke's Hospital, University of Cambridge, UK (Work address until 2007)

[2] Institut Jean le Rond d'Alembert (Paris 6), CNRS UMR 7190 Paris, France

## Abstract

Atherosclerosis is a disease characterized by the development of atherosclerotic plaques (APs) in arterial endothelium. The APs in part of an arterial wall are inhomogeneous on size and on distribution. In order to understand this in-homogeneity, the pathology of APs is analyzed by Misrepair mechanism, a mechanism proposed in our Misrepair-accumulation aging theory. **I.** In general, development of an AP is a result of repair of injured endothelium. Because of infusion and deposition of lipids beneath endothelial cells, the repair has to be achieved by altered remodeling of local endothelium. Such a repair is a Misrepair. During repair, smooth muscular cells (SMCs) are clustered and collagen fibers are produced to the lesion of endothelium for reconstructing an anchoring structure for endothelial cells and for forming a barrier to isolate the lipids. **II.** Altered remodeling (Misrepair) makes the local part of endothelium have increased damage-sensitivity and reduced repair-efficiency. Thus, this part of endothelium will have increased risk for injuries, lipid-infusion, and Misrepairs. Focalized accumulation of Misrepairs and focalized deposition of lipids result in development of a plaque. **III.** By a viscous circle between lipid-infusion and endothelium-Misrepair, growing of an AP is self-accelerating. Namely, once an AP develops, it grows in an increasing rate with time and it does not stop growing. Within part of an arterial wall, older APs grow faster than younger ones; thus old APs are always bigger than new ones, resulting in an in-homogenous distribution of APs. The oldest and the biggest AP is the most threatening one in narrowing local vessel lumen. Therefore, the self-accelerated growing of an AP is a fatal factor in atherosclerosis. **In conclusion**, development of APs is focalized and self-accelerating, because it is a result of accumulation of Misrepairs of endothelium.

## Keywords

# Introduction

Atherosclerosis is a disease characterized by the development of atherosclerotic plaques (APs) in arterial walls. Growing of an AP makes arterial wall thick and stiff, leading to the narrowness of local arterial lumen. An AP can block blood circulation when it is too big or when it drops off from the wall and becomes a blood clot. Atherosclerosis is the main cause for several fatal diseases, including coronary artery diseases, cerebral thrombosis, and aortic aneurysm. In part of an arterial wall, APs distribute in-homogenously, and they have different sizes and different shapes. Atherosclerosis is a typical aging-associated disease; however, traditional aging theories cannot interpret the in-homogeneity of APs. For example, gene-controlling theory suggests that aging is a process that is controlled completely and independently by certain genes (Fabrizio, 2010; McCormick, 2012). However, if such genes exist, they should work in the same way in all cells, and aging changes should develop homogenously. Damage-accumulation theory predicts that aging is a result of accumulation of faults (damage) (Kirkwood, 2005). However, if faults are the origins of aging, accumulation of the randomly-developed faults should result in a homogenous distribution of aging changes. In the present paper, we will demonstrate that Misrepair-accumulation aging theory is able to explain the in-homogeneity in the development of APs and other aging changes (Wang, 2009). Our discussion will tackle the following issues:

I.  Development of an atherosclerotic plaque(AP): a result of Misrepairs of endothelium

   1.1  An generalized concept of Misrepair
   1.2  Development of an AP: a result of Misrepairs for healing endothelium

II.  Development of APs: focalized and self-accelerating

   2.1   Accumulation of Misrepairs: focalized and self-accelerating
   2.2  Focalized and self-accelerated growth of APs
   2.3   Low-dose aspirin may slow down the growth of an AP

III.  Conclusions

**I.   Development of an atherosclerotic plaque (AP): a result of Misrepairs  of endothelium**

On the mechanism of development of atherosclerosis, some theories have been proposed, and they include cholesterol (lipid)-deposition theory, myoblast-mutation theory, platelet-accumulation theory, and endothelium injury-response theory (Ross, 1986). Among them, the endothelium injury-response theory is mostly accepted till today. This theory suggests that development of atherosclerosis is an inflammatory response to the injuries of endothelium. This inflammatory response is for sealing the injured endothelium, however with a result of altered remodeling of the endothelium. Thus, development of atherosclerosis can be more precisely described as a result of "Misrepair" of endothelium.

## 1.1 A generalized concept of Misrepair

To have a unified understanding of aging changes, we proposed a generalized concept of Misrepair in the Misrepair-accumulation theory. This new concept of Misrepair is defined as ***incorrect reconstruction of an injured living structure.*** This concept is applicable to all the living structures in an organism, including molecules (DNAs), cells, tissues, and organs. Misrepair of DNA is a known example. A Misrepair takes place when complete repair is impossible to achieve in the situation of a severe injury. The strategy of Misrepair is essential for maintaining the structural integrity and increasing the surviving chance of an organism in damaging environment. However, a Misrepair results in alteration of structure and reduction of functionality of a cell or a tissue. The changes of a structure caused by Misrepairs are irreversible and irremovable; thus they accumulate and deform gradually the structure, appearing as aging of it. Aging of an organism is a process of accumulation of Misrepairs of its structure. Misrepairs enable an organism to survive till reproduction age. Therefore, Misrepair mechanism is essential for the survival of a species (Wang, 2009).

## 1.2 Development of an AP: a result of Misrepairs for healing endothelium

The pathological changing of an AP with time is complicated. The changing is corresponding to a series of responses of local tissue to the injuries of endothelium. A promoting affair of this process is an injury of endothelial lining and infusion of lipids through the gap. Basement membrane is the anchoring matrix for endothelial cells; however, lipid-deposition separates endothelial cells from basement membrane. Without anchoring matrix for endothelial cells, sealing of endothelium would be impossible. For rebuilding an anchoring structure, the smooth muscular cells (SMCs) in the media layer of arterial wall proliferate, immigrate and cluster to the defect of endothelium, and produce collagen fibers. These collagen fibers are organized in a special structure and become the anchoring matrix for local endothelial cells. Infused lipids can diffuse in the space between endothelial cells and basement membrane. For restricting the diffusion of lipids, local monocytes cluster and swallow the lipids, functioning as the first barrier of lipids. However, too much lipids can overload the capacity of monocytes. Thus, a fibrotic membrane made of SMCs and collagen fibers becomes the second barrier. This barrier and basement membrane compose then a fibrotic capsule, which can isolate the lipids and the foam cells in it. Isolating the lipids into this capsule is part of the sealing of endothelium. Without this capsule, lipid-diffusion could deprive more endothelial cells from their anchoring basement membrane. SMCs can deform reversibly, and they give the capsule a certain degree of deformity.

With increase of infusion of lipids, the capsule becomes bigger and bigger by including more lipids in more layers of membrane made of SMCs and collagen fibers, and a protruding fibrotic AP develops. With death of foam cells and releasing of necrotic substances, a fibrotic AP becomes an AP. Disruption of a protruding AP can result in a big defect in endothelium, which looks like an ulcer in the arterial wall. Healing of the endothelium of this ulcer needs to undergo along the edge of the defect, resulting in a hole in the plaque. In summary, the changes of an AP in different stages are results of repairs for healing the endothelium; however because of lipid-deposition, the way of repair is changed and the structure of endothelium is altered. Development of an AP is therefore a result of accumulation of Misrepairs of endothelium.

## II. Development of APs: focalized and self-accelerating

Atherosclerotic changes occur focally, appearing as APs. But a critical question is why they develop focally rather than homogeneously in endothelium? In our view, the focalized infusion of lipids and the focalized development of APs are due to focalized accumulation of Misrepairs of endothelium. Accumulation of Misrepairs is also self-accelerating; hence the growth of APs is self-accelerating. The focalized and self-accelerated development of APs is the main reason for the in-homogeneous distribution of APs. More importantly, the self-accelerated growing of APs is a fatal factor in atherosclerosis.

### 2.1 Accumulation of Misrepairs: focalized and self-accelerating

In general, a Misrepair can increase the surviving chance of an organism from a severe injury; however it results in alteration of structure and reduction of functionality of a tissue/organ. The part of a tissue that contains a Misrepair will have reduced efficiency on making adaptive responses to environment changes; thus it becomes more sensitive to damage. In the same time, a Misrepair disturbs local substance-transportation and cell-cell communication, by which it reduces the repair-efficiency of the local tissue. With increased damage-sensitivity and reduced repair-efficiency, this part of tissue will have increased risk to be injured and to be repaired by Misrepairs. In another word, Misrepairs have a tendency to accumulate to the part of tissue and its neighborhood where an old Misrepair has taken place. Thus, accumulation of Misrepairs is focalized, resulting in the development of a spot or a plaque. The frequency of Misrepairs to this part of tissue will be increased and the affected area of tissue will be enlarged after each time of Misrepair by a viscous circle. Thus, accumulation of Misrepairs and the growing of a spot are self-accelerating. Misrepair is the origin for the development of a spot, but it is also the force to drive and accelerate the growing of the spot. Multiple Misrepairs that occur to different parts of a tissue will result in development of multiple spots. Older spots grow faster than younger ones; thus older ones are always bigger than younger ones, resulting in an inhomogeneous distribution of the spots.

### 2.2 Focalized and self-accelerated growth of APs

In atherosclerosis, lipid-infusion results in altered remodeling of endothelium. This change of endothelial structure leads to increased damage-sensitivity and reduced repair-efficiency of the local endothelium. On one hand, loss of anchoring to basement membrane makes the local endothelial cells fragile. These cells might not be able to deform efficiently for adapting to the repeated deformations of arterial wall. On the other hand, local substance-transportation and cell-cell communication can be interrupted by lipid-deposition and alteration of endothelial structure, thus the repair-efficiency of local endothelium is also reduced. Because of increased damage-sensitivity and reduced repair-efficiency, this part of endothelium has increased risk for injuries and for lipid-infusion. Each time of lipid-infusion will promote the clustering of SMCs and the production of collagen fibers for rebuilding anchoring structure for endothelial cells and for isolating the lipids. Therefore, the accumulation of lipids and the development of atherosclerotic changes are focalized, resulting in formation of plaques. In each AP, the fibrotic capsule becomes bigger and bigger with time by including more lipids in more layers of SMCs and collagen fibers. Infusion of lipids and Misrepair of endothelium enhance each other by a viscous circle; and the enlargement of a fibrotic capsule is accelerated (Figure 1). Growing of an AP is therefore self-accelerating. Namely, once an AP develops, it cannot stop growing and it grows in an accelerated rate with time. Older APs grow faster than younger ones, and older ones are always bigger than younger ones and the difference on plaque size between different plaques will be also enlarged with time (Figure 2). This results in the inhomogeneous distribution of APs.

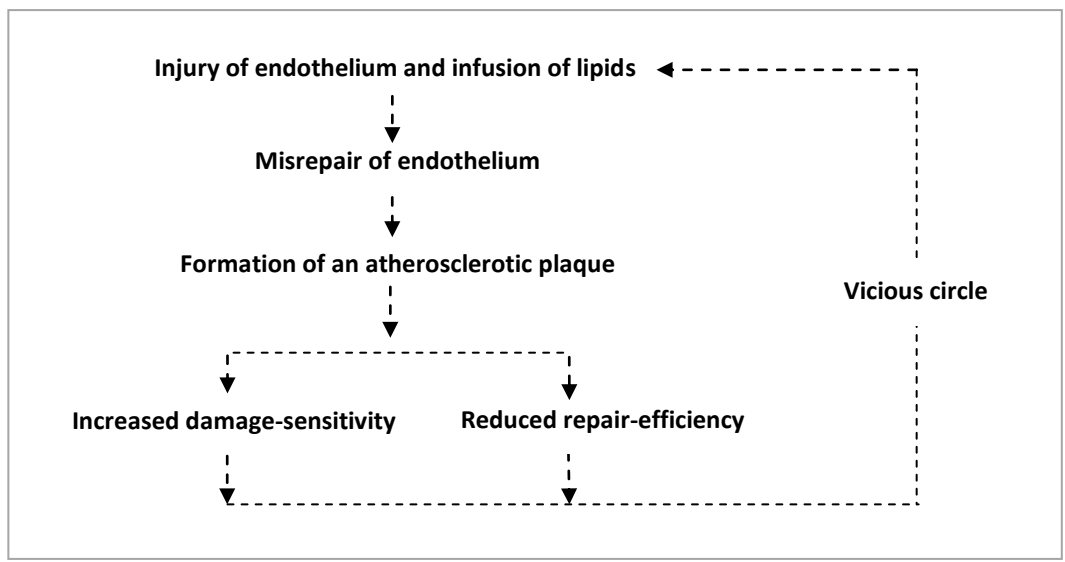

**Figure 1. A vicious circle between lipid-infusion and Misrepair of endothelium**

Atherosclerosis is promoted by an injury of endothelium and the subsequent lipid-infusion. Lipid-infusion results in Misrepair of endothelium and formation of an atherosclerotic plaque (AP). This structural change of endothelium leads to increased damage-sensitivity and reduced repair-efficacy of the local endothelium. As a result, injuries and lipid-infusion have increased opportunity to occur to the misrepaired part of endothelium. Infusion of lipids and Misrepair of endothelium compose a viscous circle, by which they enhance each other and accelerate the growth of a plaque.

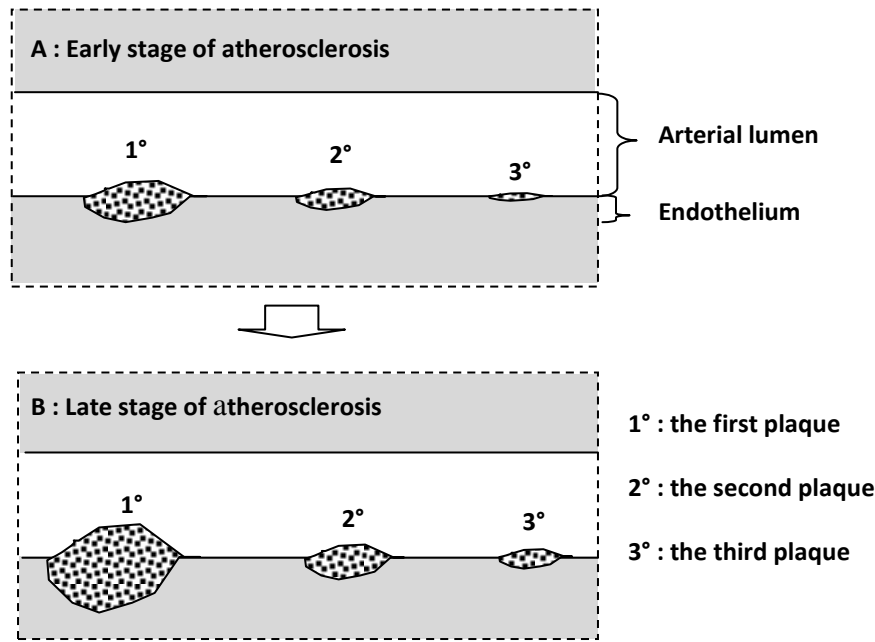

**Figure 2. Self-accelerated growth of an atherosclerotic plaque (AP)**

Development of an AP is a result of focalized accumulation of Misrepairs of endothelium. Accumulation of Misrepairs is self-accelerating, thus the development of APs is also self-accelerating. This means that older APs grow faster than younger ones. In the graph, plaque 1°, 2°, and 3° represent respectively the first-occurred, the second, and the third plaque. Since older APs grow faster than younger ones, the APs that are bigger in early stage of atherosclerosis (**A**) will be also bigger in late stage (**B**). And the difference on plaque size between different plaques will be also enlarged with time. Namely, it is permanent that the plaques are: 1° > 2° > 3° on size (**A** and **B**). The enlargement of plaque 1° is also rapider than those of plaques 2° and 3°.

## 2.3 Low-dose aspirin may slow down the growth of an AP

The size of an AP is an important factor in determining the degree of narrowing of local vessel lumen by the AP. However, for blocking an artery, the thickness of a plaque, namely the protruding degree of an AP into arterial lumen, is more critical than its size. Thus, reducing the rate of increase of the thickness of an AP could be possibly a way to delay the blockage of an artery. A strategy that can induce a flatting growth of an AP could be a solution for that. Some studies have shown that long-term application of low-dose aspirin can reduce the risk of atherosclerosis-caused heart attack (Hung, 2003; Elwood, 2006; Brotons, 2014). Aspirin has an effect of inhibiting inflammation and repair, and high-dose aspirin is dangerous by causing failure of repair. It is quite possible that low-dose aspirin may not completely inhibit repair, however makes a repair process slower. Slower repair may result in delayed healing of endothelium, prolonged infusion of lipids, and prolonged diffusion of lipids in endothelium. Prolonged diffusion of lipids between endothelial cells and basement membrane may result in the growth of an AP more along the endothelium and less into the lumen. Namely, a delayed repair can make an AP grow more flatly, and in this way the rate of increase of thickness of the AP is slowed down (Figure 3). In our view, slowing down and altering the way of growth of APs might be the mechanism why low-dose aspirin can reduce the risk of occurrence of heart attack.

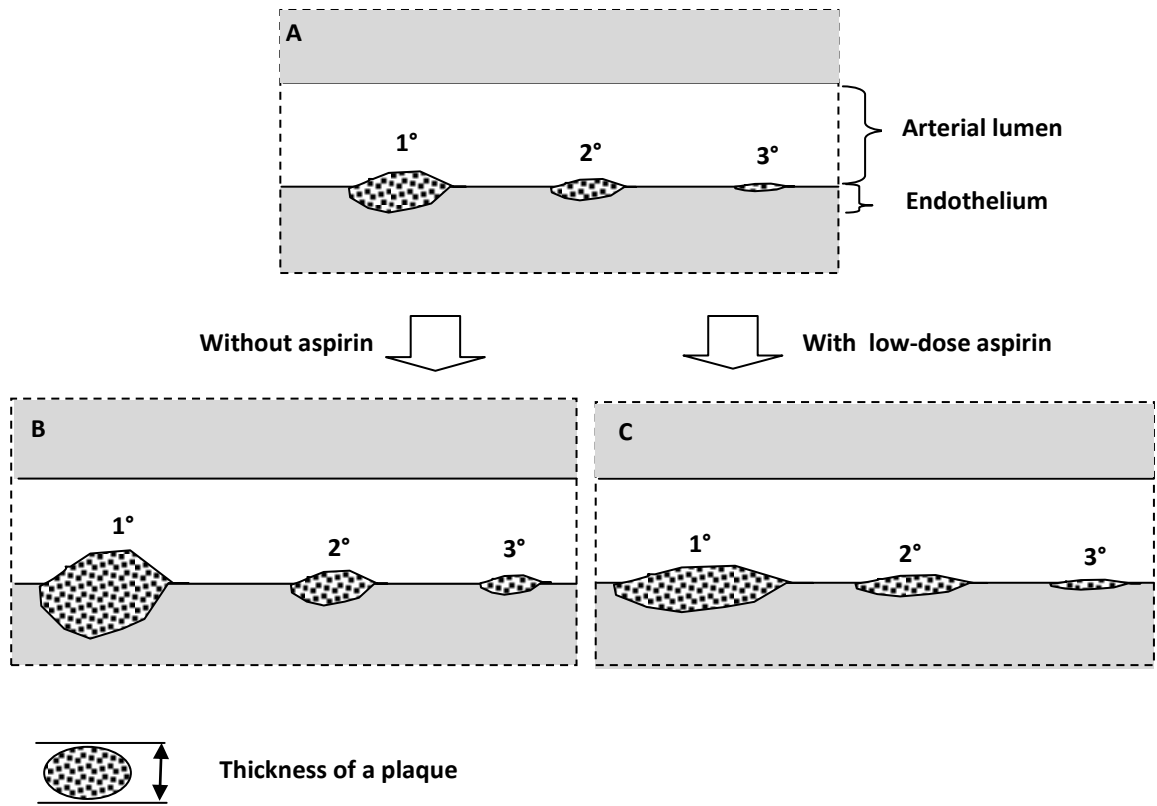

**Figure 3. Low-dose aspirin may alter the way of growth of an AP**

Low-dose aspirin may not completely inhibit repair; however it may make a repair process slower. Slower repair may result in delayed healing of endothelium, prolonged infusion of lipids, and prolonged diffusion of lipids in endothelium. Prolonged diffusion of lipids between endothelial cells and basement membrane may make the growth of an AP more along endothelium and less into lumen. In this way, an AP can grow more flatly. Plaques 1°, 2°, and 3° represent different APs in part of an arterial wall (in **A**). These APs will grow further with time. However, all the APs (1°, 2°, and 3°) in the individuals that have utilized low-dose aspirin may grow more flatly (in **C**) than the APs in those that have not used aspirin (in **B**). Slowing down the rate of increase of the thickness of APs might be the mechanism why low-dose aspirin can reduce the risk of occurrence of heart attack.

## III.   Conclusions

Development of an AP is a result of repair of injured endothelium. However, because of infusion of lipids, the repair has to be achieved by altered remodeling of the local endothelium. This is a manifestation of Misrepair. A Misrepair makes the local part of endothelium have increased damage-sensitivity and reduced repair-efficiency; thus this part of endothelium will have increased risk for injuries, for lipid-infusion, and for Misrepairs. Focalized infusion of lipids and focalized accumulation of Misrepairs result in development of a plaque. By a viscous circle between lipid-infusion and Misrepair of endothelium, growing of an AP is self-accelerating. Once an AP develops, it grows in an increasing rate with time and does not stop growing. Within part of an arterial wall, older APs grow faster than younger ones, thus old APs are always bigger than new ones, resulting in an in-homogenous distribution of APs. The oldest and the biggest AP is the most threatening one in narrowing arterial lumen. Low-dose aspirin may reduce the risk of occurrence of heart attack by slowing down and altering the way of growth of APs.

**References**

1. Brotons C, Benamouzig R, Filipiak KJ, Limmroth V, Borghi C. (2014) A Systematic Review of Aspirin in Primary Prevention: Is It Time for a New Approach? Am J Cardiovasc Drugs.
2. Elwood P, Longley M, Morgan G. (2006) My health: whose responsibility? Low-dose aspirin and older people. Expert Rev Cardiovasc Ther., 4(5), 755
3. Fabrizio P, Hoon S, Shamalnasab M, Galbani A, Wei M, Giaever G, Nislow C, Longo VD. (2010). Genome-wide screen in Saccharomyces cerevisiae identifies vacuolar protein sorting, autophagy, biosynthetic, and tRNA methylation genes involved in life span regulation. *PLoS genetics* **6**(7):e1001024.
4. Hung J. (2003) Aspirin for cardiovascular disease prevention. Medical Issues Committee of the National Heart Foundation of Australia. Med J Aust, 179(3), 147
5. Kirkwood TB. (2005). Understanding the odd science of aging. Cell 120(4): 437-47
6. McCormick M, Chen K, Ramaswamy P, and Kenyon C. (2012). New genes that extend Caenorhabditis elegans' lifespan in response to reproductive signals. Aging Cell 11(2):192-202
7. Ross R. The pathogenesis of atherosclerosis. (1986) N Engl J Med. 314:488–500.
8. Wang J, Michelitsch T, Wunderlin A, Mahadeva R. (2009) Aging as a Consequence of Misrepair – a Novel Theory of Aging. ArXiv: 0904.0575. arxiv.org